\documentclass[twocolumn]{openjournal}
\usepackage{graphicx}   
\usepackage{amsmath}    
\usepackage{amssymb,amsfonts}    
\usepackage{orcidlink}
\usepackage{verbatim}
\usepackage{longtable}
\usepackage{hyperref}

\shorttitle{I. Waisberg}
\shortauthors{I. Waisberg}

\begin{document}

\title{An SS433 flare caught by TESS \\ Fast rise times and evidence for a 13 hours quasi-period}

\newcommand{\weizmann}{Department of Particle Physics and Astrophysics, Weizmann Institute of Science, Rehovot 76100, Israel}

\email{email: idelwaisberg@gmail.com}

\author{\vspace{-1.2cm}Idel Waisberg\,\orcidlink{0000-0003-0304-743X}$^{1,2}$}

\affiliation{$^1$Independent researcher}
\affiliation{$^2$\weizmann}

\begin{abstract}
SS433 is an exotic Galactic microquasar in which there is a supercritical inflow of matter into a compact object of likely black hole nature. Here we report on an SS433 flare that occurred during a TESS observation of Sector 54 in June of 2022. The flare lasted for about 10 days and culminated in a flux increase by a factor of three relative to quiescence. The unprecedented continuous photometric coverage of the flare afforded by TESS shows that the flux can rise and fall by up to a factor of two within 4 to 7 hours. Furthermore, there is evidence that the flare peaks occur with a quasi-period $P_{\mathrm{QPO}} = 13.3$ hours. The flare can be most naturally explained by accretion rate ($\dot{M}$) variations due to oscillations in the donor star mass loss rate. We estimate $\dot{M}$ variations up to a factor of two with an increase in the photospheric radius of the accretion disk by about the same factor, implying that a transient ``common envelope'' enshrouds the binary system during the flares. If the flares are due to radial pulsations in the donor star envelope then their quasi-period $P_{\mathrm{QPO}}$ could imply a mean density $\left < \rho \right > \sim 18 \text{ kg} \text{ m}^{-3}$ for the donor star, compared to $\left < \rho_{\mathrm{RL}} \right > = 0.7-1 \text{ kg} \text{ m}^{-3}$ for a synchronized Roche-lobe filling donor star. Another SS433 lightcurve from a Sector 80 TESS observation in 2024 hints that there may be an outflow from behind the donor star. 
\end{abstract}


\section{Introduction}
\label{sec:introduction}

SS433 is a unique microquasar in which a compact object (most likely a black hole) accretes matter from a hydrogen-rich companion star at a supercritical rate of about a thousand times above the Eddington rate ($\dot M \sim 10^{-4} M_{\odot} \text{ yr}^{-1}$). This results in a plethora of outflows, most notably a pair of precessing, baryonic and semi-relativistic ($v=0.26c$) jets. The orbital lightcurve of SS433 shows three main periodicities related to the orbital motion with period $P_{\mathrm{orb}} = 13.08 \text{ days}$, in the form of eclipses of the accretion disk and donor star, and to the precession ($P_{\mathrm{prec}} \approx 163 \text{ days}$) and nutation ($P_{\mathrm{nut}} \approx 6.3 \text{ days}$) of the accretion disk and jets. SS433 is located at the center of W50, a 400 pc x 200 pc radio nebula believed to be a $\sim20,000$ yr old supernova remnant of the explosion that created the compact object. \cite{Margon84,Fabrika04,Cherepashchuk20} present comprehensive reviews of SS433.

Because the central binary is enshrouded by strong outflows and the donor star is significantly outshone by the extended accretion disk at all wavelengths, the nature of the components in SS433 are rather controversial. The nature of the donor star was thought to have finally been solved when weak absorption lines in the blue part of the spectrum with Doppler motion in phase with that expected for the donor star were detected, implying it to be an A-type giant \citep{Gies02,Hillwig08}. However, the low amplitude of the resulting radial velocity (RV) curve together with the more robust RV curve of the compact object \citep[e.g.][]{Fabrika90} implies a rather low mass ratio $q = \frac{M_X}{M_*} \sim 0.3$ and a resulting low mass $M_X \sim 3 M_{\odot}$ for the compact object \citep{Kubota10}. This is in contrast with other independent estimates of $q \gtrsim 0.8$ and $M_X \gtrsim 20 M_{\odot}$ based on the circumbinary disk \citep[e.g.][]{Blundell08,Bowler10,Bowler18} and the orbital period change \citep{Cherepashchuk19,Cherepashchuk21}. Until this contradiction is resolved the nature of the components in SS433 should remain controversial. Despite being a rather ``messy'' system, as a Galactic object that can be studied in extreme detail SS433 offers one of the best opportunities to study the supercritical accretion of matter into a (probable) black hole. 

On top of the three periodicities SS433 shows significant stochastic variability, including flares during which the optical flux increases typically by within a factor of two and more rarely by a factor of three or more \citep[e.g.][]{Fabrika03}. Several interesting behavior have been reported during SS433 flares. \cite{Blundell11} presented a comprehensive spectroscopic coverage of an SS433 flare during which the ``stationary'' H$\alpha$ emission line broadened by a factor of two and a pair of red and blueshifted emission lines at velocities of about $\pm 500 \text{ km}\text{ s}^{-1}$ appeared. The latter were originally attributed to rotation in the accretion disk but given the lack of their eclipsing they probably arose instead in outflows from the outer Lagrangian point(s) \citep{Bowler21}. Such an angular momentum-rich equatorial outflow was imaged across the Br$\gamma$ line in VLTI/GRAVITY near-infrared interferometric observations at a scale of 1 au \citep{Waisberg19} and is probably what feeds the larger scale outflows on a scale of 100 au seen in some radio images \citep{Blundell01}. Optical flares have also been linked to an increase in the speed of the baryonic jets followed by their disappearance as well as to radio flares \citep{Vermeulen93, Blundell11}. Other interesting phenomena reported during flares are the disappearance of the eclipses and a larger flux increase in (infra)red bands compared to the optical \citep{Goranskij19}. These phenomena suggest that the flares are probably related to an increase in the accretion rate onto the compact object leading to extended outflows, including significant mass loss through the outer Lagrangian point(s), that may even choke the optical jets for a few days. 

In July 2022 SS433 was located within the TESS \citep{Ricker15} field of view (Sector 54) for the first time. This provided unprecedented 
optical photometry of SS 433 when compared to previous ground-based campaigns \citep[e.g.][]{Revnivtsev04,Revnivtsev06}, with continuous photometric coverage with a 10 min cadence for about 26 days. Remarkably, a rather rare and bright SS433 flare occurred during the TESS observation. Furthermore, another SS433 TESS lightcurve was obtained in 2024 (Sector 80) during which SS433 was more quiescent but still showed rather interesting behavior. This paper is organized as follows. In Section \ref{sec:lightcurve}, we present the TESS Sector 54 lightcurve and auxiliary photometric data and discuss its main features. In Section \ref{sec:discussion}, we discuss properties of the supercritical accretion flow implied by the flare as well as its possible origin in pulsations of the donor star. Section \ref{sec:future} concludes with prospects for further SS433 flare studies. A brief discussion of the TESS Sector 80 SS433 lightcurve can be found in Apppendix \ref{app:S80}.

\section{The lightcurve}
\label{sec:lightcurve}

\begin{figure*}[]
\centering
\includegraphics[width=\textwidth]{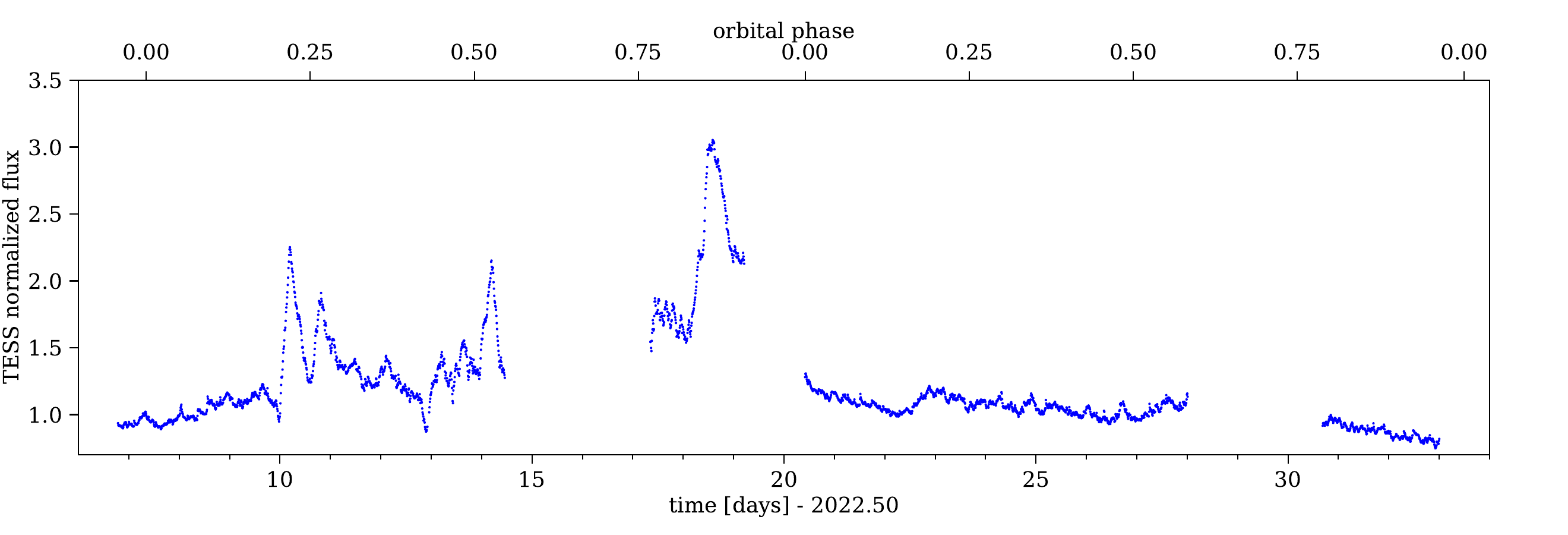} \\
\includegraphics[width=\textwidth]{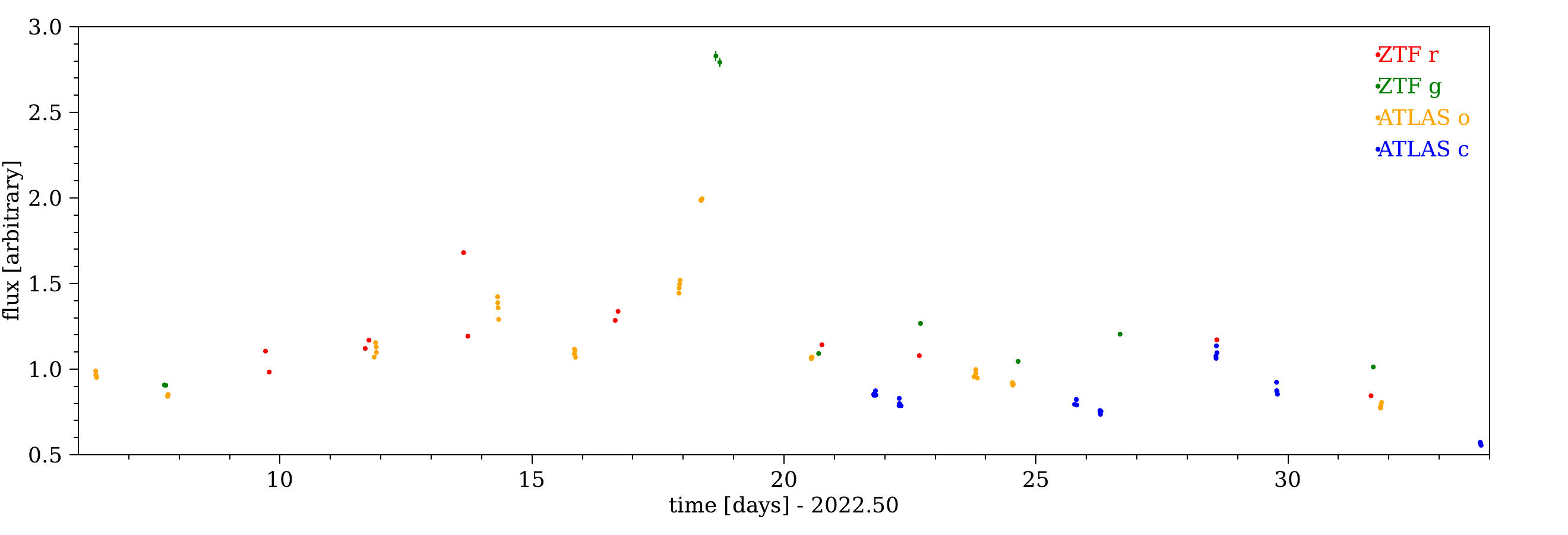}
\caption{\textbf{Top:} TESS lightcurve of SS433 in Sector 54. \textbf{Bottom:} SS433 lightcurve from all-sky surveys ZTF and ATLAS during the TESS observation. Each band is normalized by an arbitrary factor.}
\label{fig:lightcurve} 
\end{figure*}

Figure \ref{fig:lightcurve} shows the normalized SS433 TESS lightcurve from Sector 54. The data cover about 26 days (from 2022.52 to 2002.59) with a cadence of 600s but there are three gaps of about 2 days each. The TESS pixel is rather large (21") and in addition to SS433 includes a small contamination by faint sources. Details about the light curve extraction can be found in Appendix \ref{app:lightcurve}. The TESS filter is broad and extends from 6,000{\AA} to about 10,000{\AA}. Since SS433 is very red due to interstellar extinction, most of the SS433 flux caught by TESS comes from longer wavelengths. The precessional phase during the observation varied from $\psi_{\mathrm{prec}} = 0.17$ to $\psi_{\mathrm{prec}} = 0.33$ based on the precessional parameters in \cite{Cherepashchuk22}, corresponding to an accretion disk inclination varying between $i = 72^{\degr}$ and $91^{\degr}$. The data cover about two orbital periods and the orbital phases $\phi_{\mathrm{orb}}$ (where $\phi_{\mathrm{orb}}=0$ corresponds to the center of the eclipse of the accretion disk) are shown on the top x-axis in Figure \ref{fig:lightcurve} based on the orbital parameters (including the period change) reported in \cite{Cherepashchuk21}. There are no clear eclipses but this is expected given the precessional phases with the disk close to edge-on and the red wavelengths of TESS. 

\begin{figure}[]
\centering
\includegraphics[width=\columnwidth]{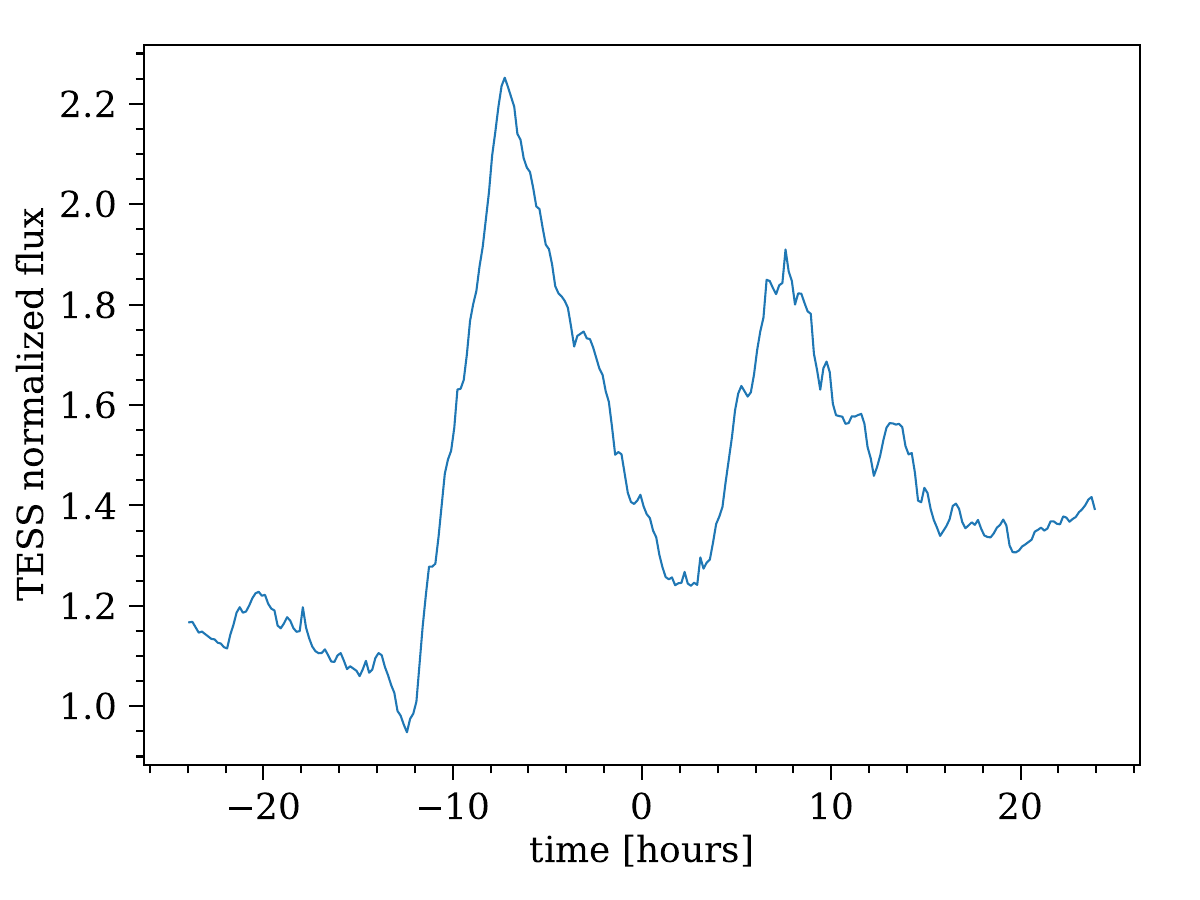}\\
\includegraphics[width=\columnwidth]{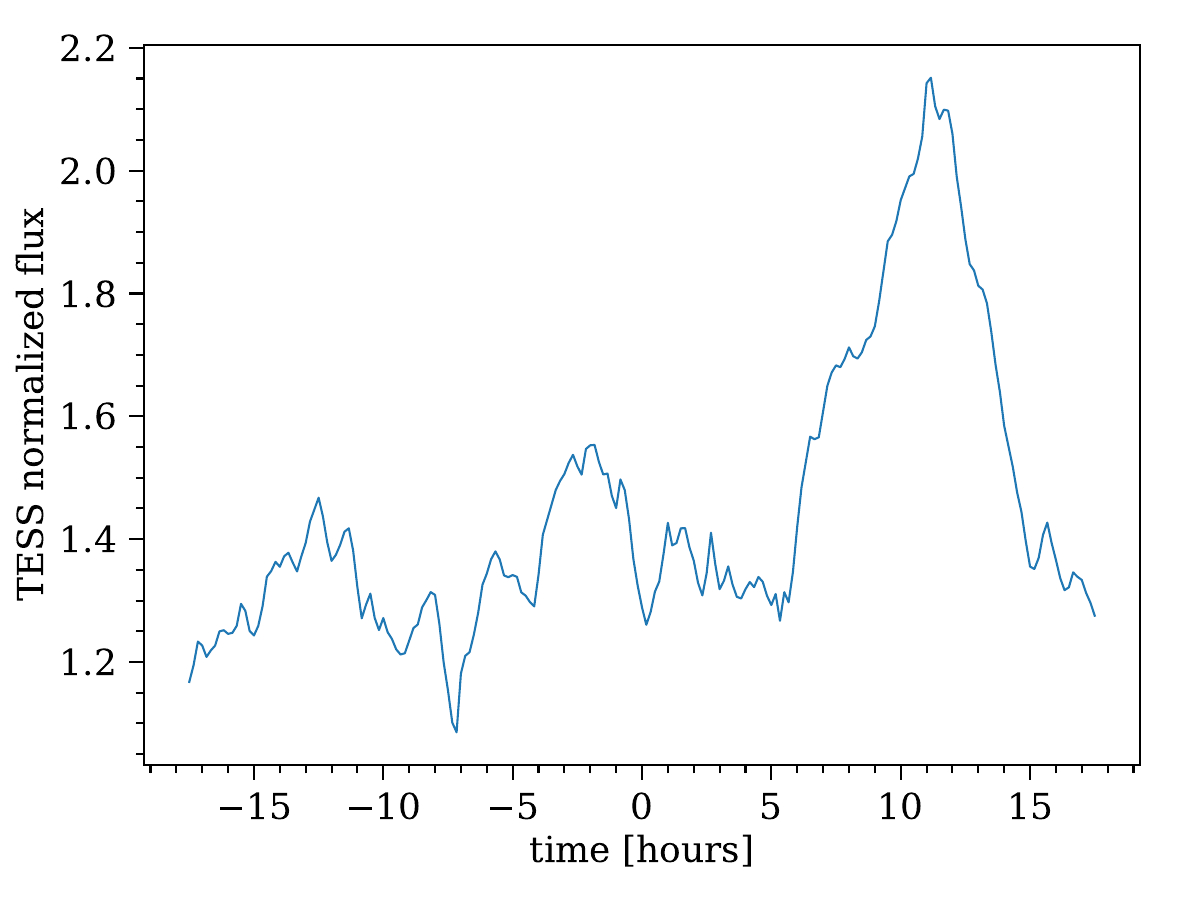} \\
\includegraphics[width=\columnwidth]{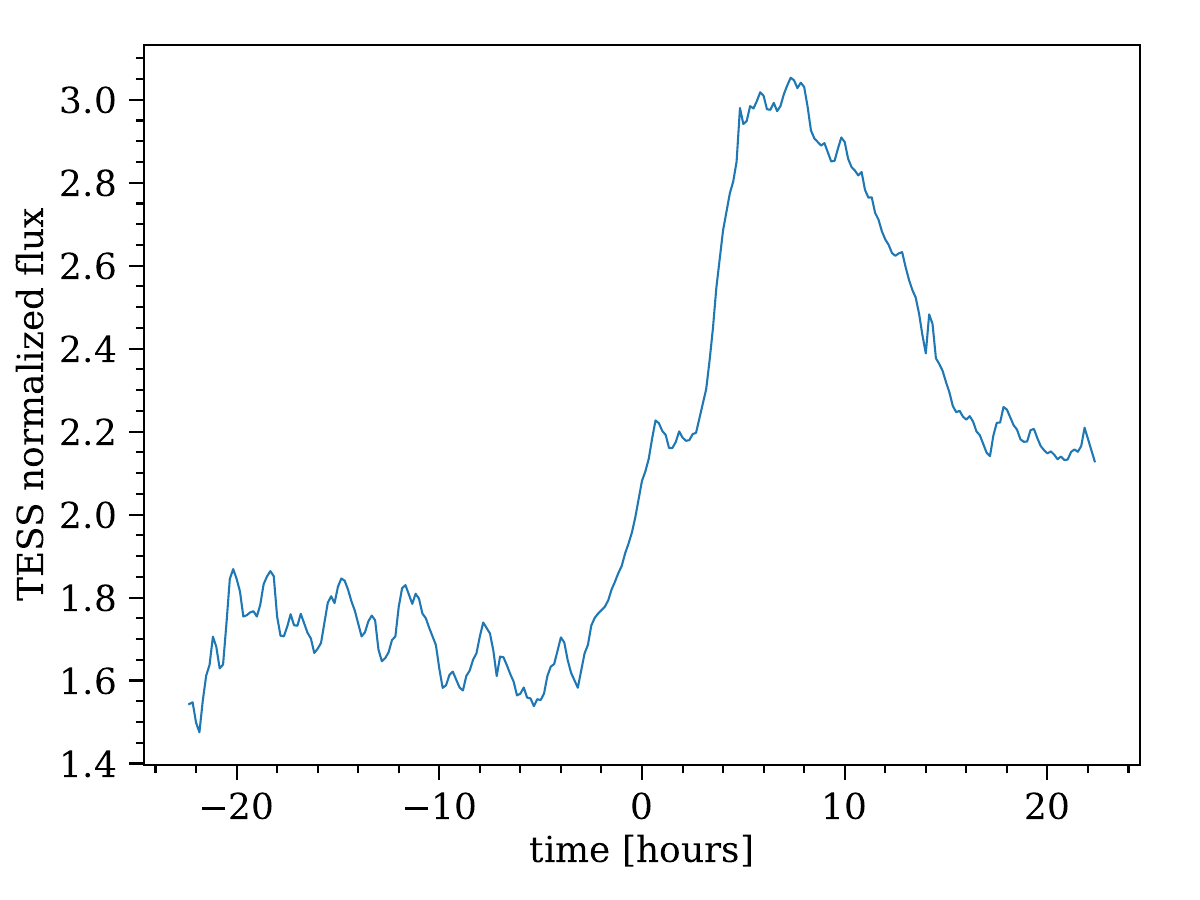}
\caption{Zoom insets of Figure \ref{fig:lightcurve} highlighting the structure of flare peaks. The figures correspond to times $9.5 < t < 11.5$, $13.0 < t < 14.5$ and $17.4 < t < 19.2$ (in the units of Figure \ref{fig:lightcurve}) respectively. The mean time has been subtracted in each figure.}
\label{fig:lightcurve_S54_zoom} 
\end{figure}

Significant flaring behavior starts at day 10 (+2022.50) and culminates in a major flare with a flux increase by a factor of about three (relative to quiescence) in day 18, after which SS433 returns to quiescence in about 2 days and remains so for the rest of the observation. Figure \ref{fig:lightcurve_S54_zoom} shows zoomed insets of Figure \ref{fig:lightcurve} highlighting the flare substructure. Within the flare peaks the TESS flux can rise (and fall back) by up to a factor of 2 within about 4 to 7 hours. 

Figure \ref{fig:lightcurve} (bottom) shows photometry from ZTF \textit{g} (4200 to 5400{\AA}) and \textit{r} (5700 to 7200{\AA}) filters, taken with 30s exposures and a camera with 1"/pixel scale, as well from ATLAS \citep{Tonry18} \textit{c} (4200 to 6500{\AA}) and \textit{o} (5600 to 8200{\AA}) filters, taken with 30s exposures and a camera with 1.86"/pixel scale, during the TESS observation. The data show the difficulty of studying SS433 flares in detail with only ground-based nightly observations. Still, the rise of the main flare peak was caught by ATLAS \textit{o} and its climax by ZTF \textit{g}, confirming without any doubt that SS433 is the origin of the flare. The flux increase in the flare climax relative to quiescence is about a factor of 2.8 in the ZTF $g$ band compared to about 3 in the TESS band, implying that SS433 became at most only slightly redder at the flare peak compared to quiescence. 

\begin{figure}[h]
\centering
\includegraphics[width=\columnwidth]{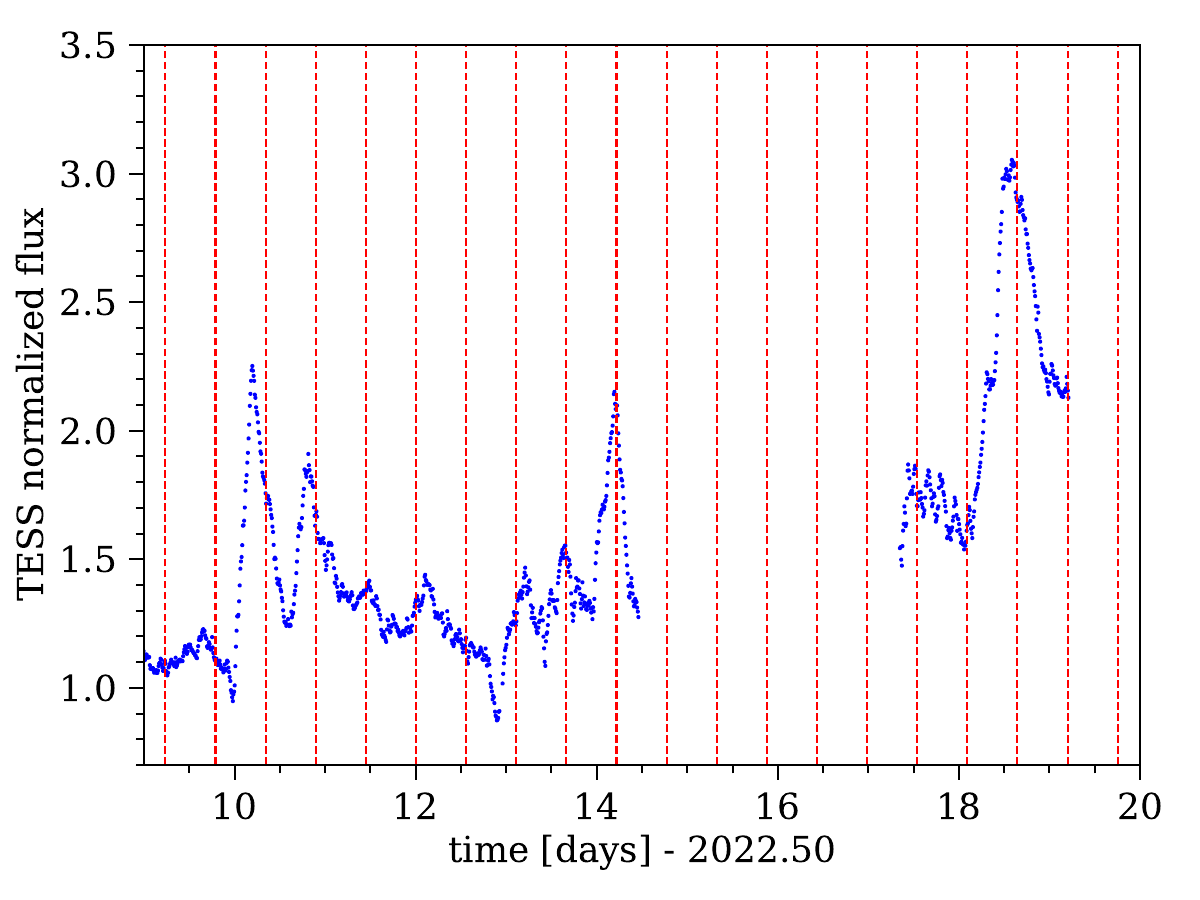} \\
\includegraphics[width=\columnwidth]{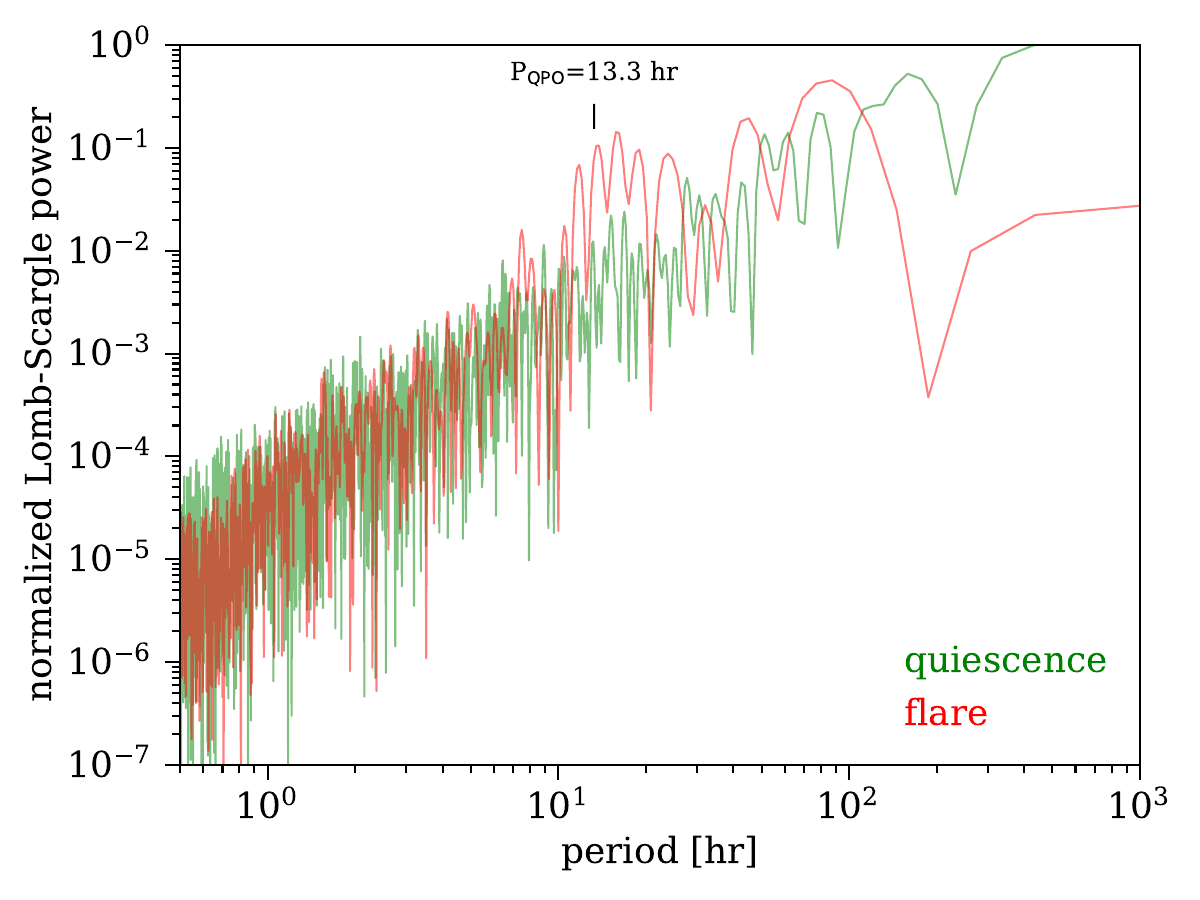} 
\caption{\textbf{Top:} Flaring portion of the SS433 TESS Sector 54 lightcurve with the times of peaks for the best-fit model with period $P_{\mathrm{QPO}} = 13.3 \text{ hr}$ marked by red dashed lines. \textbf{Bottom:} Lomb-Scargle periodogram for the SS433 TESS Sector 54 lightcurve during quiescence (green) and during the flare (red). A broad quasi-period can be seen at 13.3 hr for the latter. The quiescence curve was multiplied by 1.5 for ease of comparison.}
\label{fig:periodogram_S54} 
\end{figure}

We noticed a conspicuous repetition of flare peaks with approximately equal spacing between days 10 and 15. The sequence consists of two large peaks and two small peaks, followed by the absence of a peak for one cycle and the recovery of peaks with increasing amplitude in the next three cycles before the data break. The major peak at day 18 then occurs eight cycles later. We measured the peak times by fitting a Gaussian to each peak and then found the best-fit period. The first four peaks alone have a best fit period of 15.1 hr and the next three peaks have a best fit period of 11.7 hr. Fitting for all the peaks together we find a best fit period $P_{\mathrm{QPO}} = 13.3 \text{ hr}$. We refer to the periodicity as a quasi-period (QPO) since it is clearly not strictly periodic but still appears to be relatively well defined with a quality factor $Q \sim \frac{13 \text{ hr}}{2 \text{ hr}} = 6.5$. Figure \ref{fig:periodogram_S54} (top) shows the flaring portion of the lightcurve with the corresponding times of peaks for the best-fit model with $P_{\mathrm{QPO}} = 13.3 \text{ hr}$ in red dashed lines. 

Figure \ref{fig:periodogram_S54} (bottom) shows the Lomb-Scargle power spectral density for two data intervals: quiescence (for days 20 and later) in green and during the flare (between days 9 and 15) in red. They can be both approximately described by a power law $L(P) \propto P^{\alpha}$ as is typical for red noise with $\alpha \approx 1.6$ for the quiescent period \citep[consistent with previous ground based photometric monitoring of SS433; e.g.][]{Revnivtsev06}) and $\alpha \approx 1.8$ for the flare period. The fit for the spectral index was done for the interval $P=1-10$ hr in both cases. Furthermore, a mild and
broad peak typical of quasi-periodical oscillations can be seen at $P_{\mathrm{QPO}} = 13.3 \text{ hr}$ during the flares while it is absent during quiescence.

\section{Discussion}
\label{sec:discussion}

SS433 is ultimately powered by accretion of matter towards the compact object companion. The donor star has only a mild contribution to the total light: based on the eclipse depth $\Delta V \approx 0.6$ for the accretion disk eclipse and $\Delta V \approx 0.2$ for the donor star eclipse \citep[e.g. Figure 1 in ][]{Cherepashchuk21}, the latter contributes only about 20\% to the total optical light \citep[and should have a similarly small contribution at redder wavelengths if it is truly an A-type giant e.g.][]{Hillwig08}. Therefore, the most likely explanation for the flares is an increase in the accretion rate into the compact object leading to an extended outflow. 

The timing information provided by the unprecedented continuous coverage of an SS433 flare by TESS can be used to estimate properties of the binary system and the supercritical accretion flow. Figure \ref{fig:lightcurve_S54_zoom} shows that the TESS flux can rise (and fall back) by up to a factor of two within 4 to 7 hours. Furthermore, the quasi-period translates to a dynamical radius 

\begin{align}
R_{\mathrm{dyn}} = \left ( GM_X \left ( \frac{P_{\mathrm{QPO}}}{2 \pi} \right )^2 \right )^{1/3} = 4.3 \times 10^{11} \text{ cm } \left ( \frac{M_X}{10 M_{\odot}} \right )^{1/3}
\end{align}

\noindent where $M_X$ is the mass of the compact object. This scale is associated with the outer accretion disk (see below).

\subsection{Characteristic radii}

In order to interpret the flares we first summarize the basic scales of the binary system and the accretion disk. When relevant, we make use of the mass function estimated from the compact object radial velocity curve $\frac{M_*}{(1+q)^2} \sim 7.7 M_{\odot}$ \citep{Fabrika90}, where $M_*$ is the mass of the donor star and $q=\frac{M_X}{M_*}$ is the mass ratio\footnote{Although in principle this mass function is not immune to systematic errors (the RV of the compact object is traced by emission lines formed in its outflow), it is believed to be one of the more robust constraints in SS433.}. 

The characteristic radii of the binary system are 

\begin{enumerate}
\item The binary semi-major axis 

\begin{align}
a = \left ( G M_* (1+q) \left ( \frac{P_{\mathrm{orb}}}{2 \pi} \right )^2 \right )^{1/3} \sim 3.2 \times 10^{12} (1+q) \text{ cm} \\ = 4.2-6.4 \times 10^{12} \text{ cm for } 0.3 \leq q \leq 1 
\end{align}

\item The Roche-lobe radius of the compact object

\begin{align}
R_{L,X} = a f(q) \sim 1.2-2.4 \times 10^{12} \text{ cm for } 0.3 \leq q \leq 1 \\ 
f(q) = \frac{0.49 q^{2/3}}{0.6 q^{2/3} + \ln (1 + q^{1/3})}
\label{eq:f(q)}
\end{align}

\noindent \citep{Eggleton83}.

\item The Roche-lobe radius of the donor star 

\begin{align}
R_{L,*} = a f(1/q) \sim 2.0-2.4 \times 10^{12} \text{ cm for } 0.3 \leq q \leq 1 
\end{align}

\item The distance between the L1 point and the compact object 

\begin{align}
r_{L1} \approx a ( 0.5 + 0.227 \log_{10} q) \\ \approx 1.6-3.2 \times 10^{12} \text{ cm} \text { for } 0.3 \leq q \leq 1
\end{align}

\noindent \citep[e.g.][]{Frank02}. 

\end{enumerate}

The characteristic radii of the supercritical accretion disk are

\begin{enumerate}
\item The inner radius of the disk is of the order of the innermost stable circular orbit around the black hole

\begin{align}
R_{\mathrm{in}} = f \frac{GM_X}{c^2} = f \times 1.5 \times 10^6 \text{ cm}
\end{align}

\noindent where $f=1$ for a maximally rotating hole and $f=6$ for a non-rotating black hole. The jets are believed to be launched at a scale of a few $R_{\mathrm{in}}$. 

\item The ``spherization'' radius \citep{Shakura73} is the scale at which the accretion disk wind outflows are launched and correspond to the radius at which accreted matter reaches the Eddington luminosity 

\begin{align}
R_{\mathrm{sph}} = \frac{1}{2 \eta} \frac{GM_X}{c^2} \frac{\dot{M}}{\dot{M}_{\mathrm{Edd}}} \\ = 3.7 \times 10^9 \text{ cm} \times \left ( \frac{\eta}{0.1} \right)^{-1} \left ( \frac{M_X}{10 M_{\odot}} \right ) \left ( \frac{\frac{\dot{M}}{\dot{M}_{\mathrm{Edd}}}}{500} \right )
\end{align}

\noindent where $\eta$ is the accretion efficiency and $\dot{M}_{\mathrm{Edd}}$ the accretion rate corresponding to the Eddington luminosity ($L_{\mathrm{Edd}} = \eta \dot{M}_{\mathrm{Edd}}$). The disk is believed to become geometrically thick (aspect ratio $H/R\approx1$) at $R_{\mathrm{sph}}$. 

\item The photospheric radius $R_{\mathrm{ph}}$ corresponds to the radius in which the optical depth from outside in the direction parallel to the disk $\tau \approx 1$ (including both scattering and absorption) and for highly supercritical accretion rates can be much larger than $R_{\mathrm{sph}}$ \citep[e.g.][]{Poutanen07}. Based on the estimated luminosity and temperature of the disk from the SED, $R_{\mathrm{ph}} \sim 2 \times 10^{12} \text{ cm}$ is rather large in SS433 \citep[e.g.][]{Wagner86}. Since $R_{\mathrm{ph}} \sim R_{L,X}$, the acrretion disk is expected to fill the Roche Lobe of the compact object. This is consistent with the eclipsing lightcurves, which show that the accretion disk is extremely large (``quasi-star''). 

\item The outer radius of the accretion disk $R_{\mathrm{out}}$ would normally be somewhere between the circularization radius 

\begin{align}
R_{\mathrm{circ}} \approx a (1+q^{-1}) ( 0.5 + 0.227 \log_{10} q)^4 \\ \approx 3.8 - 8 \times 10^{11} \text{ cm} \text { for } 0.3 \leq q \leq 1
\end{align}

\noindent \citep[e.g.][]{Frank02} and the L1 point. 

\end{enumerate}

The canonical picture is that of a geometrically thin ($H/R \ll 1$) outer disk fed by a stream from the L1 point that slowly transitions to become geometrically thick ($H/R \sim 1$) at $R_{\mathrm{sph}}$. Since $R_{\mathrm{ph}} \sim R_{\mathrm{out}}$, the real disk is probably composed of an equatorial region where the mass inflow takes place surrounded by the photosphere of the outflow. 

\subsection{Supercritical accretion disk luminosity}

In order to estimate the size of the accretion rate fluctuations that cause the flares, we recall that when the mass accretion rate is highly supercritical ($\dot{M} \gg \dot{M}_{Edd}$), two competing effects become important: photon trapping and outflows. While the fraction of matter that is ultimately advected into the black hole is still an open question, canonical supercritical accretion theory predicts that the bolometric luminosity of the disk should depend only logarithmically on the mass accretion rate\footnote{The apparent luminosity (in case the observer has a favorable viewing angle) may exceed this limit substantially through geometrical beaming of the radiation by the disc funnel and the beaming factor should depend sensitively on the accretion rate. However, in the case of SS433 the minimum inclination reached by the accretion disk during the precessional and nutational cycles is about $55^{\degr}$, so that an observer on Earth does not have a view through the funnel.}

\begin{align}
L \approx L_E \left ( 1+ \ln \frac{\dot{M}}{\dot{M}_E} \right )   
\end{align}

\noindent \citep{Shakura73}. Therefore, the bolometric luminosity should be practically insensitive to $\dot{M}$ (except for gargantuan changes that are unphysical in this case) and what we are most likely seeing in the case of the SS433 flares are relatively modest accretion rate fluctuations causing the disk photosphere to increase in size and the flux density at optical/infrared wavelengths to increase at the expense of ultraviolet flux.

\subsection{The dynamics of the flare}

A plausible reason for $\dot{M}$ changes to happen are oscillations in the donor star envelope. Let us assume that an increase in $\dot{M}$ occurs in the outer accretion disk. The gas captured falls very fast towards the compact object with a viscous time $t_{\mathrm{vis}} \lesssim 0.5 \text{ day}$ as estimated from the observations of the nutation of the disk and jets \citep[see Section 6.3 in ][and references therein]{Fabrika04}\footnote{This $t_{\mathrm{vis}}$ is comparable to the free fall time at L1 and seems somewhat outrageous considering that the gas at L1 should have substantial angular momentum. However, it is observationally well established and the resulting ``impossibly high viscosity'' is one of the major unsolved issues in SS433.}. As the gas reaches $R_{\mathrm{sph}}$, it starts to get blown by the radiation pressure. The TESS flux stays roughly constant until the gas reaches the previous disk photosphere $R_{\mathrm{ph}} \sim 2 \times 10^{12} \text{ cm}$ in a time $t = R_{\mathrm{ph}}/v_{\mathrm{wind}} \sim 3 \text{ hr}$, where $v_{\mathrm{wind}} \sim 2 \times 10^3 \text{ km}\text{ s}^{-1}$ is the wind speed as estimated from the broad ``stationary'' emission lines. At this point the TESS flux starts to increase as the increased $\dot{M}$ causes the photospheric radius to expand. For an opacity dominated by electron scattering, the photospheric radius can be estimated as 

\begin{align}
\label{eq:tau}
\tau \sim \int_{R_{\mathrm{ph}}}^{\infty} n_e \sigma_T dr \sim 1 \Rightarrow R_{\mathrm{ph}} \sim \frac{\sigma_T \dot{M}}{4 \pi \mu m_p v_{\mathrm{wind}}}
\end{align}

\noindent where $\sigma_T$ is the Thomson cross section and $n_e = \frac{\dot{M}}{4 \pi r^2 \mu m_p v_{\mathrm{wind}}}$ is the electron density with $m_p$ the proton mass and $\mu$ the mean molecular weight. Therefore we take $R_{\mathrm{ph}} \propto \dot{M}$ as a rough approximation\footnote{In general $R_{\mathrm{ph}} \propto \dot{M}^{\gamma}$, where $\gamma$ includes the possible dependency of $v_{\mathrm{wind}}$ on $\dot{M}$ and other contributions to $\tau$ besides electron scattering.}. Since the luminosity of the supercritical disk is extremely weakly dependent on $\dot{M}$, it follows that the photospheric temperature $T_{\mathrm{ph}} \propto R_{\mathrm{ph}}^{-1/2} \propto \dot{M}^{-1/2}$. If the TESS band is in the Rayleigh-Jeans-like tail of the SED also during the flare, the TESS flux density $F_{\lambda} \propto R_{\mathrm{ph}}^2 T_{\mathrm{ph}} \propto \dot{M}^{3/2}$. For an $F_{\lambda}$ increasing by a factor $2-3$ relative to quiescence we therefore have that during the flare 

\begin{enumerate}
\item $\dot{M}$ increases by a factor of about $1.6-2.1$. 

\item The photospheric radius changes roughly by the same factor as $\dot{M}$. This means that at the flare climax $R_{\mathrm{ph}} \sim 4 \times 10^{12} \text{ cm} \sim a$ and the outflow enshrouds the binary system in a transient ``common envelope''. This could explain puzzling SS433 phenomena such as large-scale equatorial outflows \citep{Blundell01,Waisberg19}, ``choking'' of the baryonic jets \citep{Vermeulen93} and disappearance of eclipses \citep{Goranskij19}. It is also consistent with an increase in the duration of the primary eclipse during a flare that can be inferred from Figure 4 (middle panel) of \cite{Bowler21}. 

\item The photospheric temperature decreases by about $60-70\%$. For $T_{\mathrm{ph}} \sim 3 \times 10^4 \text{ K}$ in quiescence \citep{Wagner86} it follows that $T_{\mathrm{ph}} \sim 1.8 \times 10^4 \text{ K}$ at the flare climax. This temperature is high enough that the assumption that the TESS band is always at the Rayleigh-Jeans-like tail is justified. This is also consistent with the only very minor reddening at the flare peak between the TESS and the ZTF $g$ bands. It is also consistent with the lack of spectral index change of the continuum between 6500{\AA} and 8000{\AA} in Figure 4 (bottom panel) of \cite{Bowler21}. 

\end{enumerate}

The expected rise time of the flare peaks in this model $\frac{\Delta R_{\mathrm{ph}}}{v_{\mathrm{wind}}} \sim \frac{2 \times 10^{12} \text{ cm}}{2000 \text{ km} \text{ s}^{-1}} \sim 3 \text{ hr}$ is comparable to the measured rise times 4-7 hr. The mildly longer duration of the latter could imply that the $\dot{M}$ increase does not happen abruptly but rather over a timescale of a few hours. Even though reality is more complicated than our simple model since $\dot{M} = \dot{M} (t)$ is time-dependent, Eq. \ref{eq:tau} should still be a good approximation since $\tau$ is dominated by a small region close to the photosphere.

\subsection{Pulsations in the donor star?}

The mass accretion rate oscillations may be due to pulsations in the donor star. Since there is no 13 hr periodicity in quiescence, the pulsations would have to be transient. They could be triggered by opacity changes ($\kappa$-mechanism) induced by heating by the accretiton disk or by shocks with the acccretion disk outflow. 

There is a rather general period-mean density relation for the fundamental mode in radially pulsating stars

\begin{align}
P_{\mathrm{pul}} \left < \rho \right >^{1/2} = \mathcal{Q}
\end{align}

\noindent where $\mathcal{Q}$ is a pulsation constant \citep[e.g. Chapter 5 in ][]{Catelan15}. Its exact value depends on assumptions regarding the density $\rho$ and adiabatic constant $\gamma$ radial profiles throughout the star. Here we use the simplest form of the relation (Ritter's relation), which assumes that the density $\rho$  and the adiabatic constant $\gamma$ are constant throughout the star

\begin{align}
\label{eq:ritter}
\left < \rho \right > = \frac{3 \pi}{2 \gamma G P_{\mathrm{pul}}^2}
\end{align}

\noindent Assuming $\gamma=5/3$, this relation reproduces the mean densities of different classes of radial pulsators over eight orders of magnitude in density to within factors of a few \citep{Catelan15}. In the case of SS433, if the measured $P_{\mathrm{QPO}} = 13.3 \text{ hr}$ corresponds to the fundamental radial mode then $\left < \rho \right > \sim 18 \text{ kg} \text{ m}^{-3}$. 

The estimated $\left < \rho \right >$ can be compared to that implied by the requirement that the donor star fills its Roche lobe. This is because a Roche-lobe filling donor star has a mean density essentially determined by the orbital period and which is only very weakly dependent on the mass ratio: 

\begin{align}
\left < \rho_{\mathrm{RL}} \right > = \frac{3 \pi}{G (1+q) f^3(q) P_{\mathrm{orb}}^2}
\end{align}

\noindent where $f(q)$ is as defined in Eq. \ref{eq:f(q)}. For $P_{\mathrm{orb}} = 13.08 \text{ days}$, $\left < \rho_{\mathrm{RL}} \right > = 0.7 -1.0 \text{ kg}\text{ m}^{-3}$ for $0.3 \leq q \leq 1$. 

It is unclear how seriously the discrepancy between the two $\left < \rho \right >$ should be taken as Eq. \ref{eq:ritter} is only accurate to within a factor of a few and in the case of SS433 there are additional complications such as the fact that only one side of the donor star is heated and that the compact object also exerts a gravitational force on the envelope. On the one hand, oscillations of the donor star envelope in a Roche-lobe filling star would naturally produce $\dot{M}$ oscillations at the L1 point since $\dot{M}$ is exponentially sensitive to the overfilling factor. On the other hand, the disagreement by a factor of about twenty may hint that the canonical model of a synchronized Roche-lobe filling donor star in SS433 may need to be revised.

\subsection{Excluding alternative scenarios}

One may wonder whether it is possible that the accretion rate oscillations originate in the disk itself rather than in the donor star. We consider this to be very unlikely. The total mass in the disk $M_{\mathrm{disk}} \sim \dot{M} t_{\mathrm{vis}} \sim 10^{-7} M_{\odot}$ is not enough to sustain the flares without any increased input from the donor star. For a 50\% increase in accretion rate ($\Delta \dot{M} \sim 0.5 \times 10^{-4} M_{\odot} \text{ yr}^{-1}$), it could only sustain the flares for a time $t \sim \frac{M_{\mathrm{disk}}}{\Delta \dot{M}} \sim 9 \text{ hr}$. Furthermore, if a significant fraction of the disk mass were being consumed throughout the flare we would expect to see dips below the quiescence level in the lightcurve due to intervals with decreased $\dot{M}$; instead, the flares appear on top of the quiescent emission. Furthermore, in contrast to canonical accretion disks with very long $t_{\mathrm{vis}}$ in which mass accumulates between outbursts, in SS433 $t_{\mathrm{vis}}$ is just a few times the free-fall timescale so it is difficult to imagine a mechanism that could create periodic $\dot{M}$ oscillations in the disk itself. 

Another possibility worth considering is whether the increased outflow that creates the flares is driven directly from the donor star rather than first passing through the disk. It is certainly possible that an equatorial outflow from the donor star would cause reddening of the system and therefore an increase in the TESS flux. However, the expansion velocity of an outflow from the donor star would likely be much lower than the disk wind velocity and therefore inconsistent with the fast rise times of the flare peaks. An increased outflow from the accretion disk itself is also necessary to explain other phenomena reported in previous SS433 flares such as widening of the wind emission lines, increase in baryonic jet velocity and radio brightening of the jets \citep[e.g.][]{Blundell11}.

\section{Conclusion and future prospects}
\label{sec:future}

TESS observations in Sector 54 serendipitously caught a rare SS433 flare that culminated with a flux increase by a factor of three relative to quiescence. The observations provided unprecedented continuous coverage of the flare lasting for 10 days. This allowed for novel timing constraints such as the rise and fall times of the flare peaks of 4 to 7 hrs and most notably a quasi-periodicity of 13.3 hr in the flare peaks. 

The quasi-period matches the dynamical time in the outer acrretion disk and mass accretion rate variations due to oscillations in the donor star envelope are the most natural explanation for the flares. We have discussed implications that such a model would have for the supercritical accretion disk and flow. We found that the flares imply an increase in the accretion rate by up to a factor of two developing over a few hours and a corresponding expansion of the photospheric radius of the accretion disk by about the same factor. The resulting enshrouding of the binary system in a transient ``common envelope'' could help to explain several exotic phenomena in SS433. If the accretion rate variations are due to pulsations of the donor star, its resulting mean density is potentially in conflict with that implied by a synchronized Roche-lobe filling donor star.

Continuous and high cadence photometric observations of SS433 would be needed in order to cover other flares and determine whether the quasi-period $P_{\mathrm{QPO}} = 13.3 \text{ hr}$ is common to all the flares, to only a fraction of them or was simply a statistical anomaly of the particular flare discussed in this paper. SS433 flares are relatively common although super-flares (in which the flux increases by more than a factor of two) are somewhat rarer. As is clear from Figure \ref{fig:lightcurve} (bottom) observations with nightly cadence are not enough for a detailed study of the flares (potentially missing them altogether) and the assessment of quasi-periods on the order of $P_{\mathrm{QPO}}$. Dedicated continuous observations throughout many whole nights would be needed and either space observations or a combination of observatories on opposite sides of the globe would probably be necessary in order to avoid large 12-hr gaps. Although taxing such observations hold great promise for a better understanding of the supercritical accretion flow in SS433 and potentially for black holes in general.

\section*{Acknowledgments}
The author thanks Boaz Katz and Mordehai Milgrom for useful comments and discussion. This research has made use of the CDS Astronomical Databases SIMBAD and VIZIER \footnote{Available at http://cdsweb.u-strasbg.fr/}, NASA's Astrophysics Data System Bibliographic Services, NASA/IPAC Infrared Science Archive, which is funded by the National Aeronautics and Space Administration and operated by the California Institute of Technology, NumPy \citep{van2011numpy} and matplotlib, a Python library for publication quality graphics \citep{Hunter2007}. 
Based on observations obtained with the Samuel Oschin Telescope 48-inch and the 60-inch Telescope at the Palomar Observatory as part of the Zwicky Transient Facility project. ZTF is supported by the National Science Foundation under Grants No. AST-1440341 and AST-2034437 and a collaboration including current partners Caltech, IPAC, the Weizmann Institute for
Science, the Oskar Klein Center at Stockholm University, the University of Maryland, Deutsches Elektronen-Synchrotron and Humboldt University, the TANGO Consortium of Taiwan, the University of Wisconsin at Milwaukee, Trinity College Dublin,
Lawrence Livermore National Laboratories, IN2P3, University of Warwick, Ruhr University Bochum, Northwestern University and former partners the University of Washington, Los Alamos National Laboratories, and Lawrence Berkeley National Laboratories. Operations are conducted by COO, IPAC, and UW.

\section*{Data availability}
All the data used in this article is public and can be obtained from the respective archives.

\bibliographystyle{aasjournal}
\bibliography{main}{}

\appendix

\section{Lightcurve extraction}
\label{app:lightcurve}

\begin{figure}[h]
\centering
\includegraphics[width=0.5\columnwidth]{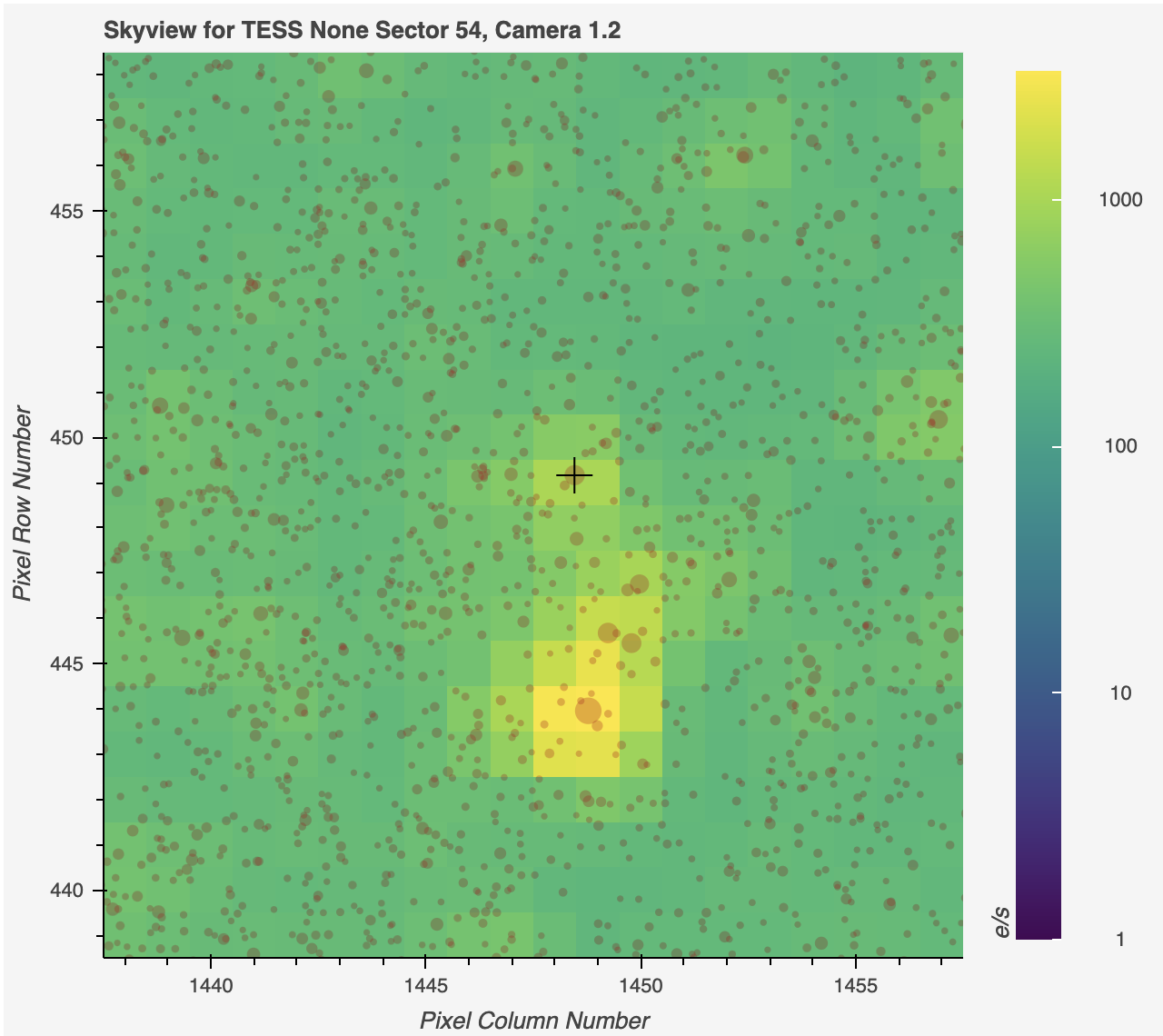} 
\caption{A cut of the TESS FFI image centered on SS433 (black cross). \textit{Gaia} sources are plotted in orange with size proportional to their brightness. The approximate sky orientation is North downwards.}
\label{fig:tess_cut} 
\end{figure}

We downloaded 20x20 pixels TESS Full Frame Image (FFI) cutouts centered on SS433 from the TESScut website \footnote{\href{https://mast.stsci.edu/tesscut}{https://mast.stsci.edu/tesscut}} \citep{Brasseur19}. Figure \ref{fig:tess_cut} shows an example TESS image with \textit{Gaia} sources marked in orange (circle size proportional to brightness) produced with the \texttt{lightkurve}\footnote{\href{https://lightkurve.github.io/lightkurve/index.html}{https://lightkurve.github.io/lightkurve/index.html}} package. The SS433 position is marked with a cross. The lightcurve was extracted by using the two bright pixels around SS433. The background was estimated from a region to the upper right of SS433 that is free from bright sources. 

\begin{figure}[h]
\centering
\includegraphics[width=0.5\columnwidth]{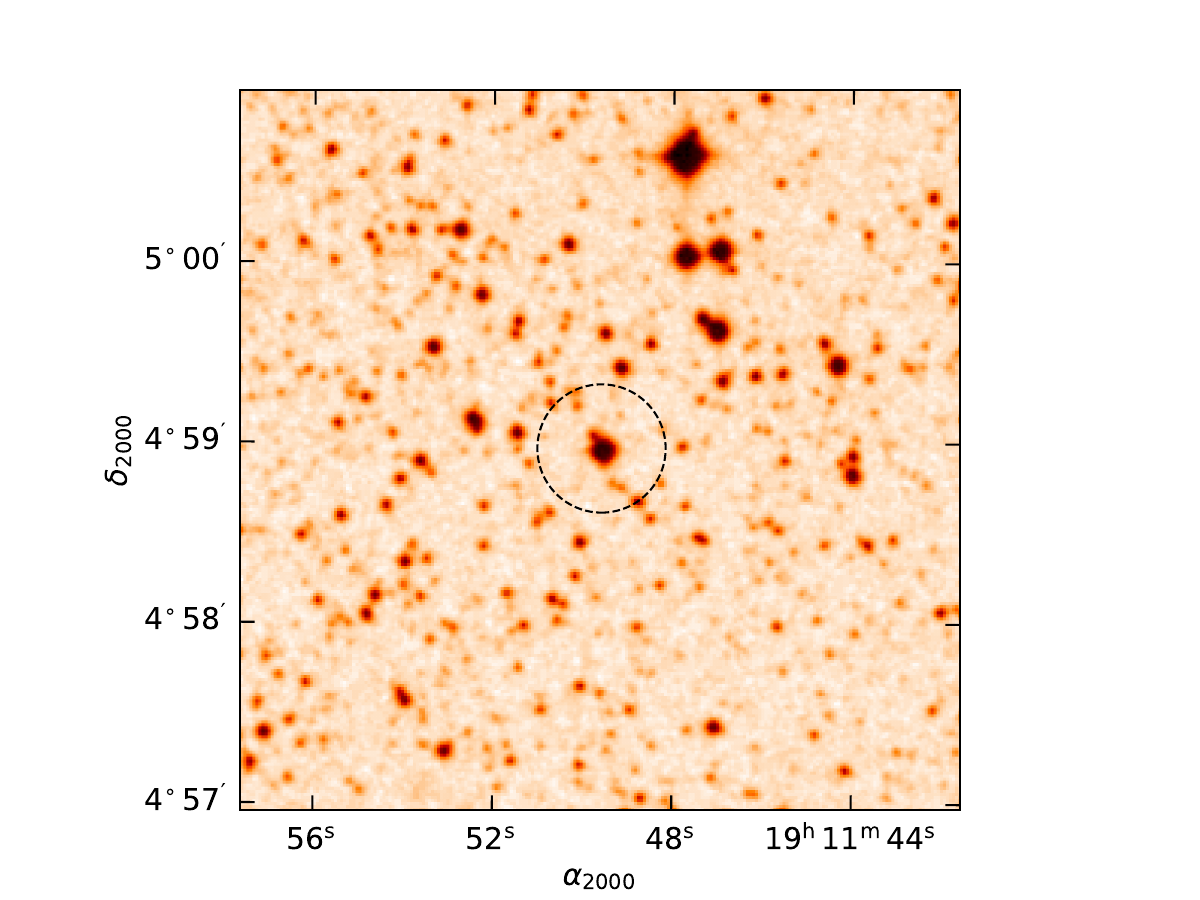} 
\caption{DSS2-Infrared image of the area around SS433. The dashed circle has a radius of 21" around SS433 and includes sources that should contribute to the TESS lightcurve.}
\label{fig:dss2_infrared} 
\end{figure}

Fig. \ref{fig:dss2_infrared} shows a 4'x4' image around SS433 from the Digitized Sky Survey 2 IR band (DSS2IR)\footnote{downloaded from the ESO Online Digitized Sky Survey: \href{https://archive.eso.org/dss/dss}{https://archive.eso.org/dss/dss}}, which has a similar wavelength coverage (6950-9000{\AA}) to TESS. A comparison with the TESS image shows that the SS433 pixels are safely free from contamination by the four bright sources forming a diamond to the northwest of SS433. However, because of the large size (21") of the TESS pixel contamination by a number of fainter sources is expected. In order to estimate the degree of contamination, we used Gaia DR3 \citep{Gaia23} photometry with the Rp filter, which has a transmission profile (6300-10000{\AA}) similar to TESS. There are 11 reported sources within a circle of radius 21" around SS433 and their total mean Rp flux amounts to 4.0\% of that of SS433. The contamination is therefore small enough to not significantly affect the SS433 lightcurve. In addition, we made use of data from the Zwicky Transient Facility \citep[ZTF; ][]{Masci19} DR17 public catalog to check that none of these sources underwent any significant variability during the TESS observations. 


\section{Sector 80 lightcurve}
\label{app:S80}

\begin{figure*}[]
\centering
\includegraphics[width=\textwidth]{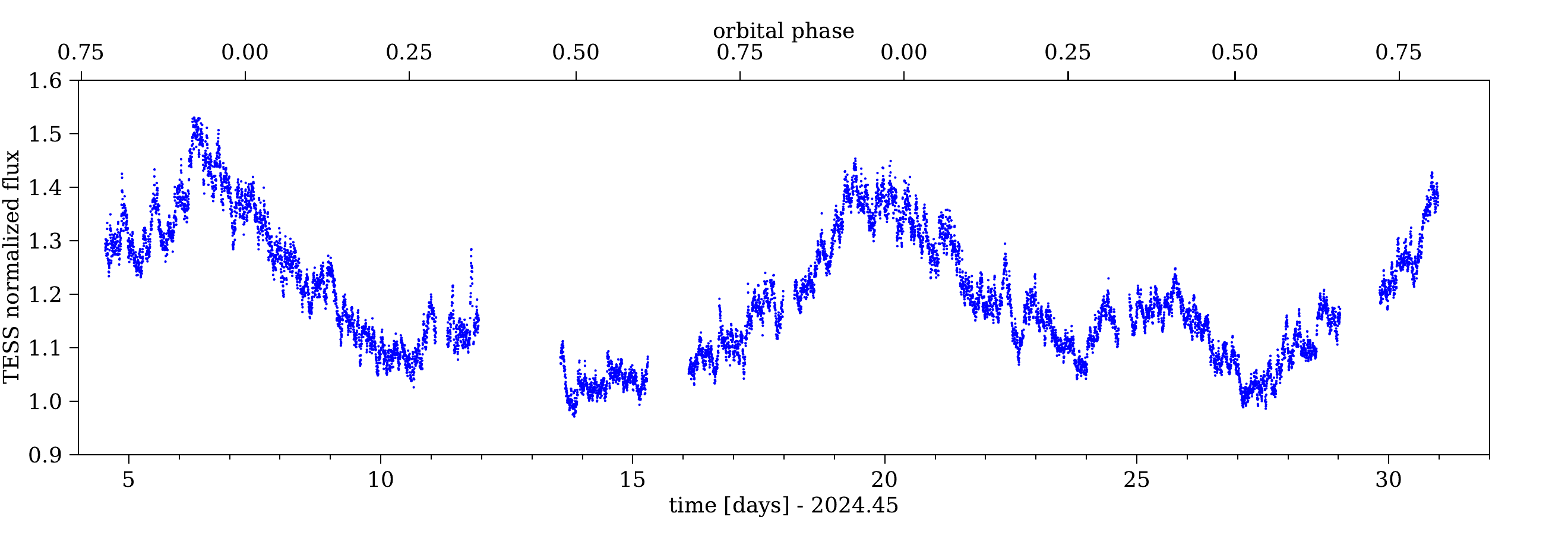} 
\caption{TESS lightcurve of SS433 during Sector 80 observations.}
\label{fig:lightcurve_S80} 
\end{figure*}

SS433 was also observed by TESS in Sector 80 for about 27 days with a cadence of 200s between 2024.46 and 2024.53, corresponding to precessional phases varying from $\psi_{\mathrm{prec}} \approx 0.58$ to $\psi_{\mathrm{prec}} \approx 0.74$ (accretion disk inclination angle varying from $i \approx 97^{\degr}$ to $i \approx 83^{\degr}$). The corresponding lightcurve is plotted in Figure \ref{fig:lightcurve_S80} with corresponding orbital phases on top. In contrast to the Sector 54 observations, SS433 was in relative quiescence with only minor ($\lesssim 10\%$) fluctuations in the timescale of a few hours. However, there are higher amplitude ($\sim 50\%$) fluctuations over a timescale of several days (we make sure that they are real as they are not observed for other stars in the same TESS FFI). Although there are only two cycles their period matches the orbital period and the maxima (minima) are centered on phase 0.0 (0.5) when the donor star (compact object) is in inferior conjunction. The fluctuations could be caused by obscuration of the system by an equatorial outflow from behind the donor star (the outflow causes reddening of the accretion disk light and therefore an increase in the TESS flux).

\begin{figure}[h]
\centering
\includegraphics[width=0.5\columnwidth]{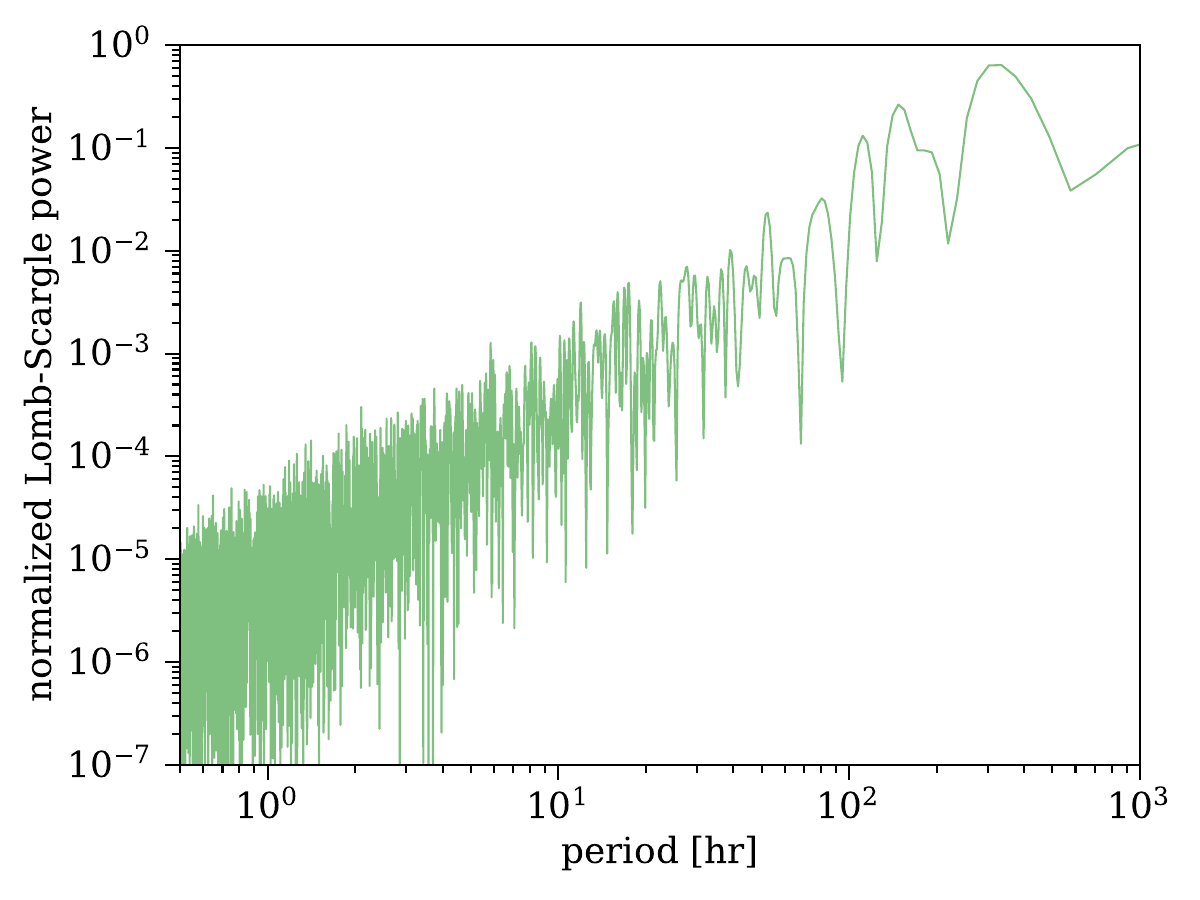} 
\caption{Lomb-Scargle periodogram for the TESS lightcurve of SS433 in Sector 80.}
\label{fig:periodogram_S80} 
\end{figure}

Figure \ref{fig:periodogram_S80} shows the Lomb-Scargle periodogram for the Sector 80 SS433 lightcurve. It has a typical red noise shape with $\alpha=1.35$ ($L \propto P^{\alpha}$) fitted between for $P=1-10$ hr without any evidence for the quasi-periodicity observed during the S54 flare. 

\end{document}